\documentclass[manuscript]{acmart}

\usepackage{enumitem}

\AtBeginDocument{%
  \providecommand\BibTeX{{%
    \normalfont B\kern-0.5em{\scshape i\kern-0.25em b}\kern-0.8em\TeX}}}

\copyrightyear{2023} 
\acmYear{2023} 
\setcopyright{acmlicensed}
\acmConference[RecSys '23]{Seventeenth ACM Conference on Recommender Systems}{September 18--22, 2023}{Singapore, Singapore}
\acmBooktitle{Seventeenth ACM Conference on Recommender Systems (RecSys '23), September 18--22, 2023, Singapore, Singapore}
\acmPrice{15.00}
\acmDOI{10.1145/3604915.3608832}
\acmISBN{979-8-4007-0241-9/23/09}

\begin{document}

\title[]{Adaptive Collaborative Filtering with Personalized Time Decay Functions for Financial Product Recommendation}

\author{Ashraf Ghiye}
\orcid{0009-0008-7088-0337}
\affiliation{%
  \institution{BNP Paribas Corporate and Institutional Banking}
  \department{Global Markets, Data \& AI Lab}
  \city{Paris}
  \country{France}
}

\affiliation{%
  \institution{École Polytechnique}
  \department{Computer Science Laboratory, LIX}
  \city{Palaiseau}
  \country{France}
}


\author{Baptiste Barreau}
\orcid{0000-0001-9045-0141}
\affiliation{%
  \institution{BNP Paribas Corporate and Institutional Banking}
  \department{Global Markets, Data \& AI Lab}
  \city{Paris}
  \country{France}
}

\author{Laurent Carlier}
\orcid{0009-0000-0631-6100}
\affiliation{%
  \institution{BNP Paribas Corporate and Institutional Banking}
  \department{Global Markets, Data \& AI Lab}
  \city{Paris}
  \country{France}
}

\author{Michalis Vazirgiannis}
\orcid{0000-0001-5923-4440}
\affiliation{%
  \institution{École Polytechnique}
  \department{Computer Science Laboratory, LIX}
  \city{Palaiseau}
  \country{France}
}

\authorsaddresses{Correspondence:
\href{mailto:ashraf.ghiye@bnpparibas.com}{ashraf.ghiye@bnpparibas.com}}


\begin{abstract}

Classical recommender systems often assume that historical data are stationary and fail to account for the dynamic nature of user preferences, limiting their ability to provide reliable recommendations in time-sensitive settings. This assumption is particularly problematic in finance, where financial products exhibit continuous changes in valuations, leading to frequent shifts in client interests. These evolving interests, summarized in the past client-product interactions, see their utility fade over time with a degree that might differ from one client to another. 
To address this challenge, we propose a time-dependent collaborative filtering algorithm that can adaptively discount distant client-product interactions using personalized decay functions. Our approach is designed to handle the non-stationarity of financial data and produce reliable recommendations by modeling the dynamic collaborative signals between clients and products. We evaluate our method using a proprietary dataset from BNP Paribas and demonstrate significant improvements over state-of-the-art benchmarks from relevant literature. Our findings emphasize the importance of incorporating time explicitly in the model to enhance the accuracy of financial product recommendation.

\end{abstract}






\maketitle
\renewcommand{\shortauthors}{A. Ghiye et al.}

\section{Introduction}

In many e-commerce applications, such as music streaming services, the primary goal of recommender systems is to maximize user engagement. Products in these applications typically have low maintenance costs and negligible inventory risk. On the contrary, corporate and institutional banks operating as market makers need to ensure liquidity in financial markets by constantly offering bid (buy) and ask (sell) prices which involve taking frequent positions in both directions. The client can send an electronic notification asking for detailed information, which shows her interest in buying or selling a specific product, through a process known as a request for quotation (RFQ). Conversely, salespeople may contact clients to offer competitive quotes on products held by the bank. Having a reliable recommendation algorithm for corporate banks has two main consequences: (1) increasing customer satisfaction by providing them with relevant investment opportunities and (2) reducing inventory risk by enabling risk managers and salespeople to be more proactive in anticipating the needs of their clients.

Modeling the clients’ behaviour in terms of their historical interests is of paramount importance to understand their profile and anticipate their needs. Some clients adopt a more conservative approach, displaying enduring interest in specific products; others adopt a rather diverse strategy, showing limited recurring interest in products. To provide personalized recommendations that resonate with all clients (e.g., to match their risk appetite), it is crucial to consider the context of their previous trades. For instance, products similar to those a client had inquired a few days ago may be more salient in determining her current interest compared to products she consulted months ago. 

To that end, we present a time-dependent recommender system tailored to the unique characteristics of the financial industry. We start by defining the clients' and products' context in terms of their first-hop neighbourhoods in the user-item interaction graph. In this work, we assume that interest persists for a certain period after the interactions have occurred, after which the utility of the interactions progressively fades over time. Therefore, we propose to personalize the rate at which this decay occurs. We conduct an extensive study on a proprietary database of G10 Bonds RFQs to test the effectiveness of different aggregation functions in encoding the dynamic collaborative signals. In addition, we compare our proposed method to relevant benchmarks from prior studies (HCF~\cite{barreau_history-augmented_2020}, LightGCN~\cite{LightGCN}). Our findings highlight the importance of incorporating time explicitly in the modeling process. In summary, our contributions are:

\begin{itemize}
    \item   We propose personalized time decay functions to aggregate the past, assuming that the utility of past interactions decreases monotonically with time.
    \item 	We demonstrate that using time explicitly to down-weight user-item interactions enhances the performance of graph collaborative filtering algorithms.
    \item	Our approach shows substantial improvements compared with different aggregation strategies, from a time-agnostic average to more advanced methods like Gated Recurrent Unit~\cite{gru} and Attention~\cite{attention}. 
\end{itemize}

In the following sections, the terms "user-item" or "client-product" will be used interchangeably to align with the vocabulary commonly used in the recommender systems literature.

\section{Background and Related Work}

The main objective of a recommender system is to predict the level of interest a user $u$ exhibits towards a specific item $i$ by scoring any user-item pair from the set of all users $U$ and all items $I$. Typically, historical interest are stored in the form of an interaction matrix, denoted by $\textbf{A}$, where each entry $a_{ui}$ corresponds to the feedback (e.g., ratings for explicit or clicks for implicit) provided by a user $u$ for an item $i$. In our study, we deal with an implicit feedback problem, where interactions correspond to a binary signal, i.e., the requests for quotation (RFQs).

Collaborative Filtering (CF) is a widely used technique in recommender systems~\cite{yohan}, with the core task of predicting the missing entries in the interaction matrix $\textbf{A} \in \{0,1\}^{\mid U \mid \times \mid I \mid}$. For instance, Matrix Factorization (MF)~\cite{mf, bpr} approximates $\textbf{A}$ by the product of two lower-rank matrices: $\textbf{P} \in \mathbb{R}^{\mid U \mid \times E}$ and $\textbf{Q} \in \mathbb{R}^{\mid I \mid \times E}$, such that $\hat{\textbf{A}} = \textbf{P} \times \textbf{Q}^\top$ recovers the original matrix with minimal loss of information.
Each row $\textbf{p}_u$ in $\textbf{P}$ and each row $\textbf{q}_i$ in $\textbf{Q}$ are commonly referred to as user and item embeddings, respectively, with $E$ representing the embedding size. Recent neural recommender models~\cite{ncf} use similar embedding components but improve the interaction modeling by using a stack of non-linear transformations to learn better representations and incorporating side information to improve the factorization~\cite{xue_deep_2017}.

\subsection*{Related Work}

Conventional CF algorithms fail to account for crucial contextual factors, such as the time or order of user interactions. Prior research has attempted to incorporate time into these methods through various approaches~\cite{tars_sysrev, campos_tars}, such as adding time as a third dimension in the rating matrix~\cite{koren_bias, temporal_cf, dtmf_2020}, using time in the pre-processing step to down-weight item ratings based on recency~\cite{exponential_2022, ding_time}, or only considering items rated within a fixed time window~\cite{timewindow}. However, most of these approaches have focused on explicit feedback and received less attention in the context of implicit feedback.

Another line of research has used historical interactions as pre-existing features to introduce dynamics in neural network recommender systems~\cite{svdplus, deepmfyoutube, att_hist}. In particular, Sequential Recommender Systems (SRS) model the patterns in user history to provide dynamic recommendations~\cite{survery_SRSs, SRSs, TIASA_sequential_2019}. One such method is Caser~\cite{caser}, which uses convolutional filters to model short- and long-term behaviors in a user's history. However, SRS models often focus on modeling only one side of the interactions, namely the user history.

More recently, Graph Neural Networks (GNNs) have emerged as the state-of-the-art approach for recommender systems~\cite{pinsage, hamilton_graph}. By representing historical user-item interactions as a bipartite graph, GNNs capture collaborative signals by learning representations of users and items based on their connectivity. Specifically, GNNs refine a user's (or an item's) embedding by aggregating the embeddings of their local neighbourhoods~\cite{LightGCN, wang_neural_2019}. However, most existing methods represent interactions as a static graph and fail to record the order of the user-item pairs, limiting their ability to capture short-term preferences. Few studies have begun to explore this challenge~\cite{dgcf, sr_gnn}. Notably, Wu et al.~\cite{GNNBook2022} demonstrate how to construct dynamic graphs to frame the next-item prediction task as a link prediction problem.

Our research targets financial recommender systems and focuses on the link prediction approach. Building on the previous work of Barreau et al.~\cite{barreau_history-augmented_2020}, we introduce time in the propagation step to discount the utility of past interactions.

\section{Methodology}

We propose a new temporal aggregation function that aims at producing better time-aware recommendations by adaptively down-weighting the utility of past interactions - we call it Adaptive Collaborative Filtering (ACF).

Our goal is to learn a mapping function to generate temporal embedding for each user $u$ (and item $i$), at any given time $t$, based on their sequential behaviour in the past. The model prediction at time $t$, measured in days, is defined as the inner product of user and item final embeddings at that time: $\hat{a}_{ui}^t = \sigma (\langle \textbf{p}_u^t, \; \textbf{q}_i^t \rangle)$, where $\sigma$ is the sigmoid function. The model's outputs are used as the ranking scores to generate daily recommendations.

We decompose the temporal embedding of user $u$ at time $t$ into two components: (1) a static embedding, denoted by $\textbf{e}_u^t = f_1(\textbf{x}_u^t)$, which captures the user's features (i.e., IDs), and (2) a history embedding, denoted by $\textbf{h}_u^t = \text{AGG}(\{ \textbf{e}_j^{t_j}: t_j <t\})$, which aggregates the embeddings of the historical items, where $\textbf{e}_j^{t_j} = f_2(\textbf{x}_j^{t_j})$ represents the static embedding of the $j^\text{th}$ item inquired by the user before $t$.

The final user embedding is computed using a $1\times1$ convolutional network $\phi_1$. The filters operate along the embedding axis, and considers both static and history embeddings as separate channels (Barreau et al.~\cite{barreau_history-augmented_2020}): 

\begin{equation}
\textbf{p}_u^t = \phi_1(\textbf{e}_u^t, \; \textbf{h}_u^t)
\end{equation}

Similarly, we apply the same decomposition on the item side to obtain the final item embedding using $\phi_2$:

\begin{equation}
\textbf{p}_i^t = \phi_2(\textbf{e}_i^t, \; \textbf{h}_i^t)
\end{equation}

The sub-scripted functions $(f_1, \phi_1)$ and $(f_2, \phi_2)$ denote separate weights for the user and item sides, respectively.

Our research focuses on the dynamic component. We define the temporal neighbourhood of user $u$ at query time $t$ as the set $\mathcal{N}_t(u) = [(i_1, t_1), (i_2, t_2), \dots , (i_n, t_n)]$, which consists of the $n$ most recent \textbf{time-stamped} interactions of user $u$. $t_j$ describes the timestamp when the interaction occurred ($t_j < t$). The temporal neighbourhood of an item $i$ is defined similarly. The history embedding is obtained by mapping the temporal set into a vector in $\mathbb{R}^{E}$.

Barreau et al.~\cite{barreau_history-augmented_2020} propose taking a simple average of the embeddings over the set of history, which can be viewed as linearly propagating the static embeddings on the user-item interaction graph~(LightGCN\cite{LightGCN}):

\begin{equation}
    \label{eq:mean_agg}
    \textbf{h}_u^t = \frac{1}{\mid \mathcal{N}_t(u) \mid} \sum_{(j, t_j) \in \mathcal{N}_t(u)} \textbf{e}_j^{t_j}; \; \quad \textbf{h}_i^t = \frac{1}{\mid \mathcal{N}_t(i) \mid} \sum_{(j, t_j) \in \mathcal{N}_t(i)} \textbf{e}_j^{t_j}
\end{equation}

where $| \mathcal{N}_t(u) |$ and $| \mathcal{N}_t(i) |$ denote the neighbourhood size. The current formulation assigns equal importance to all past interactions, regardless of their temporal order. However, we suppose that recent RFQs provide a stronger signal of interest than older ones. Consequently, it is crucial to incorporate temporal information in the aggregation to capture the evolving nature of user preferences and account for the shifting dynamics of user-item interactions.

Moreover, not all clients share the same level of trading activity. Some may have multiple inquiries per day, while others may have long periods between interactions. Considering a fixed number of past interactions can lead to sets of inquiries with varying time intervals and possibly starting from different points in time. As a result, relying solely on time to discount older interactions may be sub-optimal in representing the diverse range of time horizons. 

To address the above issues, we propose to learn different time factors to reflect the relative importance of an item $i$ to a user $u$ given the time the interaction has occurred $t_j$. More generally, we want to learn personalized weighting coefficients $w_{ui}^{t_j}$ to model the varying utility of the interaction based on time and the user-item pair:

\begin{equation}
    \label{eq:timedecay}
    \begin{aligned}
    \textbf{h}_u^t = \sum_{(j,t_j) \in \mathcal{N}_t(u)} w_{uj}^{t_j} \; \textbf{e}_j^{t_j}
    \end{aligned}  
\end{equation}

To that end, we propose to parameterize the aggregation function with power law kernels and personalized decay rates, i.e., we model the weights $w_{ui}^t$ to be inversely proportional to the time elapsed since the interaction ($\Delta t_j = t - t_j$):

\begin{equation}
    \label{eq:aggdecay}
    w_{uj}^{t_j} = \frac{1}{\Delta t_j ^{\alpha_{ui}}}
\end{equation}

where $\alpha_{ui}$ are learnable decay rates given by a feed-forward neural network. To compute the decay rate, we pass the concatenated user and item's embeddings through a linear layer and apply a sigmoid activation to constrain $\alpha_{ui}$ in the range $[0,1]$ - as it has shown better empirical results compared to other activation functions (e.g., ReLU):

\begin{equation}
    \label{eq:decayfactor}
    \alpha_{uj} = \sigma(g_1(\textbf{e}_u \; || \; \textbf{e}_j^{t_j})) = \sigma(\textbf{W}_1(\textbf{e}_u \; || \; \textbf{e}_j^{t_j}))
\end{equation}

\begin{figure}[!h] 
    \centerline{
    \includegraphics[scale=0.65, width=0.5\textwidth]{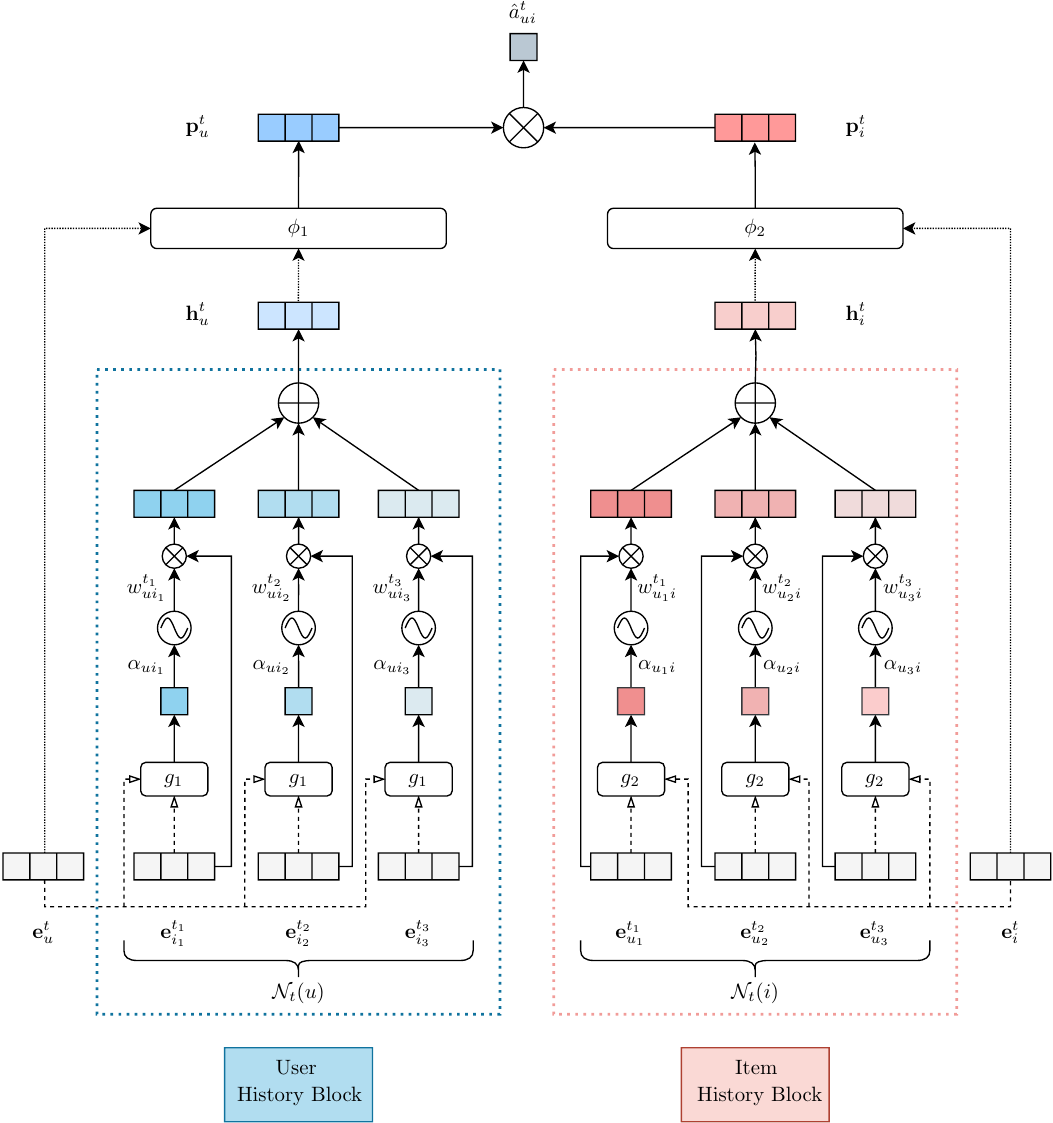}}
    \caption{Summary of the model architecture for a history of size $n=3$. The static embedding of the $k^{\text{th}}$ item (user) in the user's (item's) history is denoted by $\textbf{e}_{i_k}^{t_k}$ ($\textbf{e}_{u_k}^{t_k}$). Dashed and dotted arrows correspond to the concatenate and stack operations, respectively. $\phi_1$ and $\phi_2$ combine the ID embedding (static embedding) and first-order signal (history embedding). The history embedding is obtained by taking a weighted sum of the static embeddings in the temporal neighbourhood. The weights $w_{ui}^t$ are optimized to capture the varying utility of the interaction given the user-item pair, and the time elapsed since the interaction occurred.}
    \centering
    \label{fig:model}
\end{figure}

Figure~\ref{fig:model} illustrates the global architecture of the model.

\section{Experiments}
 
In this study, we conducted a series of experiments to evaluate the performance of our proposed method (\textbf{ACF}) against two competing benchmarks: \textbf{HCF}~\cite{barreau_history-augmented_2020} and \textbf{LightGCN}~\cite{LightGCN}. In addition, we compared against two other standard benchmarks: a vanilla matrix factorization (\textbf{MF}) and a most popular rule (\textbf{POP}). Finally, we implemented variants of \textbf{ACF} to study the impact of different aggregation functions on the model performance. All models are described below:

\begin{itemize}[leftmargin=*]
    \item \textbf{POP:} We recommend the items (or users) based on their popularity in the training set. 

    \item \textbf{MF:} We maintain the same learning procedure while excluding the dynamic component from the model, i.e., the history blocks. The only learnable parameters are the two lookup tables $f_1$ and $f_2$ of combined size $(\mid U \mid + \mid I \mid ) \times E$.

    \item \textbf{HCF:} We use an average to aggregate the history (Eq.~\ref{eq:mean_agg}). $\phi_1$ and $\phi_2$ are each a 2-layers $1 \times 1$ convolutional network with 64 and 32 filters respectively. The total number of trainable parameters is $(\mid U \mid + \mid I \mid ) \times E$ + $2 \times (2 \times 64 \times 32)$.

    \item \textbf{LightGCN:} We use a one-layer LightGCN, which is similar to HCF with the only difference of using a linear combination for the final embedding, i.e., $\textbf{p}_u^t = \textbf{e}_u^t + \frac{1}{2}\textbf{h}_u^t$. This algorithm has the same number of parameters as MF.

    \item \textbf{ACF-LP, ACF-LE:} We keep the same architecture as HCF, with extra $2 \times (2 \times E)$ trainable parameters to produce the learnable decay rates (Eq.~\ref{eq:decayfactor}). \textbf{P} and \textbf{E} indicate respectively a power law or an exponential kernel for the decay.

    \item \textbf{ACF-P, ACF-E:} Same as previous models, but with fixed decay rates ($\alpha_{ui}=1)$.

    \item \textbf{ACF-ATT:} The weights of Equation~\ref{eq:timedecay} are defined as attention coefficients directly: $w_{ui}^{t_j} = \sigma(\textbf{W}(\textbf{e}_u^t \; || \; \textbf{e}_j^{t_j} + \textbf{e}_{\Delta t_j}))$, where $ \textbf{e}_{\Delta t_j}$ is a time embedding given by a look-up table ($f_3$) of size $100 \times E$. 
 
    \item \textbf{ACF-GRU:} We use a Gated Recurrent Unit (GRU) with one hidden layer of size $E$. We take the last hidden state of the sequence to represent the history embedding.

\end{itemize}

\subsection{Data}
We use a proprietary dataset from BNP Paribas, which consists of 492,702 daily requests for quotation (RFQs) between 2118 institutional clients and 2246 governmental bonds from the G10 countries recorded over 453 days. In addition to their IDs, each client (resp. bond) at time $t$ is represented by the list of their $n=20$ previous bonds (resp. clients) within a window of the past 100 days. If this list is empty, we replace the history embedding with a zero vector embedding.

To obtain three contiguous sets, $D_{\text{train}}, D_{\text{val}}$ and $D_{\text{test}}$, we apply a straightforward temporal split to avoid data leakage. The training set encompasses one year, from $01/08/2018$ to $31/07/2019$; the validation set spans one month, from $01/08/2019$ to $31/08/2019$; the test set also spans one month, from $01/09/2019$ to $30/09/2019$.

\subsection{Training}

We train all models using a symmetrized version of the \textit{Bayesian Personalized Ranking} (BPR) loss~\cite{bpr}, defined as follows:

\begin{equation}
    \label{eq:bpr_loss}
    \mathcal{L}_{BPR} = - (1-p) \sum_{(t,u,i,j)\in D_u} \log(\sigma(\hat{a}_{ui}^t - \hat{a}_{uj}^t)) -p \sum_{(t,u,v,i)\in D_i} \log(\sigma(\hat{a}_{ui}^t - \hat{a}_{vi}^t))
\end{equation}

\sloppy where $p$, the probability of sampling a negative client, is originally set to zero. We set it to $0.5$ to give equal importance to both tasks, i.e., recommending clients and products. Let $D=\{(t,u,i) \; | \; a_{ui}^t=1\}$ be the set of all observable interactions. Then, $D_u = \{(t,u,i,j) \; | \; a_{ui}^t=1, a_{uj}^t\neq 1\}$ is obtained by considering all possible quadruplets $(t,u,i,j)$, such that the triplet $(t,u,i) \in D$, but $(t,u,j) \notin D$. The objective is to learn a personalized ranking, such that each user defines a ranking relation between items. In this case, the user $u$ is more interested in item $i$ than item $j$ at time $t$. We also define $D_i = \{(t,u,v,i) \; | \; a_{ui}^t=1, a_{vi}^t\neq 1\}$ similarly. Finally, we approximate the loss in equation~\ref{eq:bpr_loss} via negative sampling.

Our method employs non-positive uniform sampling with time constraints. Specifically, for every positive sample $(t, u, i)$, we uniformly sample a valid negative item $j$ for $D_u$ if it satisfies two conditions: it is not in the temporal neighbourhood of user $u$ at time $t$ ($j \notin \mathcal{N}_t(u)$), and if the bond has not expired yet. Similarly, a negative user $v$ is considered valid for $D_i$ if $v \notin \mathcal{N}_t(i)$, and if the client is still registered in the catalogue.

To ensure a fair comparison, we train multiple instances of each model with different hyper-parameters. We select the best model based on the lowest validation loss. For training, we use Adam~\cite{kingma2014adam} with early stopping to prevent over-fitting. The final evaluation in the subsequent section is performed on the test set.

\subsection{Evaluation}

After training, the model is used to generate daily recommendations. For each day $t$, the model produces embeddings for all users and items. The evaluation is carried out on the scoring perimeter, defined as the Cartesian product of all valid users and items on that given day, which encompasses all catalogued clients and unexpired bonds.

Consider a query $q$ that can be either a user $u$ or an item $i$, for which we aim to provide a list of recommended items or users, respectively. Let $R_q = [r_{q}^{(1)}, \dots, r_{q}^{(l)}, \dots, r_{q}^{(k)}]$ denote the recommendation list in descending order of output scores, where $l$ denotes the \textit{rank} position, and $k$ represents the total number of items or users. To measure the quality of recommendations, we use two standard metrics~\cite{ir_eval}:

\begin{itemize}
    \item Mean Reciprocal Rank (MRR) defined as:
\end{itemize}
    
\begin{equation}
    \label{eq:mrr}
    \text{MRR}_t = \frac{1}{\mid Q_t \mid} \; \sum_{q \in Q_t} \frac{1}{\textit{rank}(r_{q}^{(f)})}
\end{equation}

\begin{itemize}
    \item Mean Average Precision (mAP) defined as:
\end{itemize}
    
\begin{equation}
    \label{eq:map}
    \text{mAP}_t = \frac{1}{\mid Q_t \mid} \; \sum_{q \in Q_t} \text{AP}(q)
\end{equation}

where $r_{q}^{(f)}$ refers to the first relevant recommendation in the list, and $\text{AP}(p)$ is the average precision. $Q_t$ represents the set of queries on day $t$, which consists of either the clients who made inquiries or the products that were inquired on that day. The two query sets lead to user-side and item-side scores for each metric. To evaluate the model's effectiveness in providing relevant recommendations for both clients and risk managers, we use the symmetrized MRR (s-MRR) and mAP (s-mAP) scores~\cite{barreau_history-augmented_2020}, defined as the harmonic mean of user- and item- MRR and mAP, respectively.

\section{Results}

Table~\ref{tab:metrics} reports the average performance of all models during the test period. The results demonstrate that our proposed method (ACF) outperforms all other baselines: The best model, ACF-LE, shows substantial improvements in both metrics - with gains of up to $16.86\%$ in s-mAP and $12.65\%$ in s-MRR compared to HCF. ACF-LP and ACF-ATT follow with second and third-best scores, with only ACF-GRU underperforming the benchmark. One possible interpretation is that recurrent-based models are not well-suited to handle sequences with variable time intervals between actions~\cite{rnn_fail}.

Moreover, The performance gap between HCF and LightGCN can be attributed to the choice of an appropriate combination layer for the final embeddings. Indeed, HCF combines the static and history embeddings using two stacks of convolutional layers, which proves to be a more effective strategy than LightGCN's approach. In turn, both models outperform MF, which disregards the history blocks. It suggests that the collaborative signal carried in the first-order connectivity of the graph is valuable for the recommendation task.

\begin{figure*}[!b]
    \centering
    \scalebox{0.83}{\includegraphics[scale=0.2, width=1\textwidth]{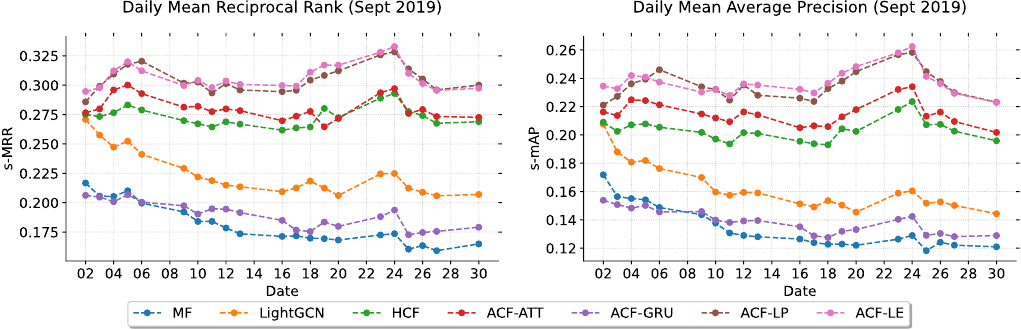}}
    \caption{Daily evolution of the evaluation metrics during the test period, shown for (left) s-MRR and (right) s-mAP.}
    \label{fig:rolling_metrics}
\end{figure*}

The daily evaluation shown in Figure~\ref{fig:rolling_metrics} reveals that most graph-based models maintain a relatively stable performance throughout the test period, except for ACF-GRU and LightGCN. These models exhibit a decrease in performance as we move away from the training period, showing their inability to cope with evolving interests, possibly due to the inappropriate aggregation module (GRU) or the ineffective combination layer (LightGCN). As a result, their performance follows the same trend as MF, which only uses static embeddings as input. The history blocks allow the model to update the latent representations by incorporating new incoming interactions at inference time. As a result, models equipped with history blocks can maintain a high level of performance over prolonged periods, requiring less frequent retraining. Notably, our proposed time-encoding strategy consistently outscoring all other baselines, demonstrating its superiority in providing reliable recommendations that capture evolving interests in dynamic settings.

\begin{table}[htbp]
\centering

\caption{Symmetrized MRR (s-MRR) and mAP (s-mAP) are reported in percentage. The Gain row shows the percentage improvement of the best model (\textbf{highlighted in bold}) compared to the benchmark (\underline{underlined}).}
\label{tab:metrics}
\scalebox{0.9}{\begin{tabular}{|c|cc|}
\hline
\multicolumn{1}{|c|}{Model / Metric} & s-MRR (\%) & s-mAP (\%) \\ \hline
\textbf{POP} & 1.08 & 0.64 \\
\textbf{MF} & 17.78 & 13.18 \\
\hline
\textbf{LightGCN} & 22.13 & 16.01 \\
\textbf{HCF} &  \underline{27.28} &  \underline{20.35} \\
\hline
\textbf{ACF-GRU} & 18.80 & 13.72 \\
\textbf{ACF-ATT} & 28.28 & 21.54 \\
\textbf{ACF-LP} &  30.54 &  23.59 \\
\textbf{ACF-LE} & \textbf{30.73} & \textbf{23.78}\\
\hline
\ Gain (\%) & 12.65 & 16.86 \\ \hline
\end{tabular}}
\end{table}

To analyze the main components of our model, we compare the impact of different types of aggregation and neighbourhood sizes on the model's performance in Table.~\ref{tab:ablation_metrics}. We summarize the following observations:

\begin{itemize}
    \item Larger neighbourhood sizes consistently correlate with better performance, as observed by the general trend of increasing s-MRR and s-mAP values as $n$ increases. Interestingly, even when using a small history size (e.g. $n=5$), any model with a history block surpasses MF, suggesting that using the first-order interactions to enrich the embeddings boosts the performance of collaborative filtering.
     
    \item The choice of a time kernel is crucial. Our experimental results show that an exponential kernel is highly sensitive to decay rates, as seen in the stark difference between ACF-LE and ACF-E. Conversely, a power law with fixed decay performs nearly as good as personalized decays, as shown by the slight difference between ACF-LP and ACF-P. One possible explanation could be that interest decay might be relatively uniform across users and items in our dataset and can be adequately modeled with a power law.
    
    \item Our experiments show that time-decayed approaches outperform attention mechanisms with time embeddings, although both perform better than a simple average (HCF). The performance gap between the two can be attributed to the following reasons: (1) the history sequences exhibit significant variability with many gaps and missing values, which can hinder the effectiveness of attention mechanisms; (2) our dataset is limited in size, whereas these models require large datasets to capture the full range of patterns and dependencies. Moreover, attention is typically employed to highlight the most relevant interactions in a sequence, whereas our approach assumes all history to be important but with decreasing utility over time.

\end{itemize}

The results of this section are twofold. First, incorporating a history block in recommendation models to leverage the first-order interactions can enhance collaborative filtering.
Second, selecting suitable aggregation functions, such as time-decayed functions, is crucial for capturing the varying utility of historical data and achieving optimal performance.

\begin{table*}[!h]
\centering
\caption{Evaluation Metrics for different aggregation functions (vertically) and history sizes $n$ (horizontally). For $n=0$, all models collapse to an MF and hence share the same performance. The best model is \textbf{highlighted in bold}}
\label{tab:ablation_metrics}
\scalebox{0.9}{\begin{tabular}{|c|ccccc|ccccc|}
\hline
\multicolumn{1}{|c|}{} & \multicolumn{5}{c|}{s-MRR (\%)} & \multicolumn{5}{c|}{s-mAP (\%)} \\ \cline{2-11} 
\multicolumn{1}{|c|}{Model / History size} & $n=0$ & $n=5$ & $n=10$ & $n=15$ & $n=20$ & $n=0$ & $n=5$ & $n=10$ & $n=15$ & $n=20$ \\ \hline
\textbf{HCF} & 17.78 & 19.98 & 23.14 & 24 & 27.28 & 13.18 & 14.52 & 17.22 & 18.05 & 20.35\\
\hline
\textbf{ACF-GRU} & 17.78 & 17.39 & 18.12 & 18.24 & 18.80 & 13.18 & 12.69 & 13.28 & 13.47 & 13.72\\
\textbf{ACF-ATT} & 17.78 & 22.72 & 24.65 & 25.97 & 28.28 & 13.18 & 17.32 & 18.11 & 19.68 & 21.54 \\
\hline
\textbf{ACF-P} & 17.78 & 24.29 & 26.01 & 27.55 & 30.24 & 13.18 & 18.5 & 19.97 & 21.37 & 23.49\\
\textbf{ACF-E} & 17.78 & 18.32 & 20.40 & 21.83 & 28.04 & 13.18 & 13.76 & 15.11 & 16.40 & 21.54 \\
\hline
\textbf{ACF-LP} & 17.78 & \textbf{24.85} & \textbf{26.39} & \textbf{28.43} & 30.54 & 13.18 & \textbf{19} & \textbf{20.37} & \textbf{21.93} & 23.59\\
\textbf{ACF-LE} & 17.78 & 23.78 & 25.56 & 27.18 & \textbf{30.73} & 13.18 & 18.01 & 19.59 & 20.58 & \textbf{23.78} \\
\hline
\end{tabular}}
\end{table*}

\section{Conclusion}

In this paper, we introduce ACF, a novel time-dependent collaborative filtering algorithm for capturing the dynamic signal of user-item interactions. We argue that personalizing the varying utility of the interactions using time is crucial for recommendation tasks. Furthermore, experimental results show the effectiveness of time-decayed approaches compared to more complicated aggregations, such as attention and recurrent-based models. This work represents an initial attempt to integrate time explicitly into the propagation step of dynamic graphs. In future work, we plan to investigate more design choices for time-aware propagation with higher-order connectivity and extend our approach by incorporating dynamic attributed graphs, which involve nodes whose features change over time.

\bibliographystyle{ACM-Reference-Format}
\bibliography{acf-base}


\begin{thebibliography}{33}


\ifx \showCODEN    \undefined \def \showCODEN     #1{\unskip}     \fi
\ifx \showDOI      \undefined \def \showDOI       #1{#1}\fi
\ifx \showISBNx    \undefined \def \showISBNx     #1{\unskip}     \fi
\ifx \showISBNxiii \undefined \def \showISBNxiii  #1{\unskip}     \fi
\ifx \showISSN     \undefined \def \showISSN      #1{\unskip}     \fi
\ifx \showLCCN     \undefined \def \showLCCN      #1{\unskip}     \fi
\ifx \shownote     \undefined \def \shownote      #1{#1}          \fi
\ifx \showarticletitle \undefined \def \showarticletitle #1{#1}   \fi
\ifx \showURL      \undefined \def \showURL       {\relax}        \fi
\providecommand\bibfield[2]{#2}
\providecommand\bibinfo[2]{#2}
\providecommand\natexlab[1]{#1}
\providecommand\showeprint[2][]{arXiv:#2}

\bibitem[Barreau and Carlier(2020)]%
        {barreau_history-augmented_2020}
\bibfield{author}{\bibinfo{person}{Baptiste Barreau} {and}
  \bibinfo{person}{Laurent Carlier}.} \bibinfo{year}{2020}\natexlab{}.
\newblock \showarticletitle{History-{Augmented} {Collaborative} {Filtering} for
  {Financial} {Recommendations}}. In \bibinfo{booktitle}{\emph{Fourteenth {ACM}
  {Conference} on {Recommender} {Systems}}}. \bibinfo{publisher}{ACM},
  \bibinfo{address}{Virtual Event Brazil}, \bibinfo{pages}{492--497}.
\newblock
\showISBNx{978-1-4503-7583-2}
\urldef\tempurl%
\url{https://doi.org/10.1145/3383313.3412206}
\showDOI{\tempurl}


\bibitem[B\"{u}ttcher et~al\mbox{.}(2010)]%
        {ir_eval}
\bibfield{author}{\bibinfo{person}{Stefan B\"{u}ttcher},
  \bibinfo{person}{Charles Clarke}, {and} \bibinfo{person}{Gordon~V. Cormack}.}
  \bibinfo{year}{2010}\natexlab{}.
\newblock \bibinfo{booktitle}{\emph{Information Retrieval: Implementing and
  Evaluating Search Engines}}.
\newblock \bibinfo{publisher}{The MIT Press}.
\newblock
\showISBNx{0262026511}


\bibitem[Campos et~al\mbox{.}(2014)]%
        {campos_tars}
\bibfield{author}{\bibinfo{person}{Pedro~G. Campos}, \bibinfo{person}{Fernando
  Díez}, {and} \bibinfo{person}{Iván Cantador}.}
  \bibinfo{year}{2014}\natexlab{}.
\newblock \showarticletitle{Time-aware recommender systems: a comprehensive
  survey and analysis of existing evaluation protocols}.
\newblock \bibinfo{journal}{\emph{User Modeling and User-Adapted Interaction}}
  \bibinfo{volume}{24}, \bibinfo{number}{1} (\bibinfo{date}{Feb.}
  \bibinfo{year}{2014}), \bibinfo{pages}{67--119}.
\newblock
\showISSN{1573-1391}
\urldef\tempurl%
\url{https://doi.org/10.1007/s11257-012-9136-x}
\showDOI{\tempurl}


\bibitem[Cho et~al\mbox{.}(2014)]%
        {gru}
\bibfield{author}{\bibinfo{person}{Kyunghyun Cho}, \bibinfo{person}{Bart van
  Merri{\"e}nboer}, \bibinfo{person}{Caglar Gulcehre}, \bibinfo{person}{Dzmitry
  Bahdanau}, \bibinfo{person}{Fethi Bougares}, \bibinfo{person}{Holger
  Schwenk}, {and} \bibinfo{person}{Yoshua Bengio}.}
  \bibinfo{year}{2014}\natexlab{}.
\newblock \showarticletitle{Learning Phrase Representations using {RNN}
  Encoder{--}Decoder for Statistical Machine Translation}. In
  \bibinfo{booktitle}{\emph{Proceedings of the 2014 Conference on Empirical
  Methods in Natural Language Processing ({EMNLP})}}.
  \bibinfo{publisher}{Association for Computational Linguistics},
  \bibinfo{address}{Doha, Qatar}, \bibinfo{pages}{1724--1734}.
\newblock
\urldef\tempurl%
\url{https://doi.org/10.3115/v1/D14-1179}
\showDOI{\tempurl}


\bibitem[Covington et~al\mbox{.}(2016)]%
        {deepmfyoutube}
\bibfield{author}{\bibinfo{person}{Paul Covington}, \bibinfo{person}{Jay
  Adams}, {and} \bibinfo{person}{Emre Sargin}.}
  \bibinfo{year}{2016}\natexlab{}.
\newblock \showarticletitle{Deep Neural Networks for YouTube Recommendations}.
  In \bibinfo{booktitle}{\emph{Proceedings of the 10th ACM Conference on
  Recommender Systems}} (Boston, Massachusetts, USA)
  \emph{(\bibinfo{series}{RecSys '16})}. \bibinfo{publisher}{Association for
  Computing Machinery}, \bibinfo{address}{New York, NY, USA},
  \bibinfo{pages}{191–198}.
\newblock
\showISBNx{9781450340359}
\urldef\tempurl%
\url{https://doi.org/10.1145/2959100.2959190}
\showDOI{\tempurl}


\bibitem[de~Borba et~al\mbox{.}(2017)]%
        {tars_sysrev}
\bibfield{author}{\bibinfo{person}{Eduardo~Jos{\'e} de Borba},
  \bibinfo{person}{Isabela Gasparini}, {and} \bibinfo{person}{Daniel
  Lichtnow}.} \bibinfo{year}{2017}\natexlab{}.
\newblock \showarticletitle{Time-Aware Recommender Systems: A Systematic
  Mapping}. In \bibinfo{booktitle}{\emph{Human-Computer Interaction.
  Interaction Contexts}}, \bibfield{editor}{\bibinfo{person}{Masaaki Kurosu}}
  (Ed.). \bibinfo{publisher}{Springer International Publishing},
  \bibinfo{address}{Cham}, \bibinfo{pages}{464--479}.
\newblock
\showISBNx{978-3-319-58077-7}


\bibitem[Ding and Li(2005)]%
        {ding_time}
\bibfield{author}{\bibinfo{person}{Yi Ding} {and} \bibinfo{person}{Xue Li}.}
  \bibinfo{year}{2005}\natexlab{}.
\newblock \showarticletitle{Time weight collaborative filtering}.
  \bibinfo{publisher}{ACM}, \bibinfo{address}{Bremen Germany},
  \bibinfo{pages}{485--492}.
\newblock
\showISBNx{978-1-59593-140-5}
\urldef\tempurl%
\url{https://doi.org/10.1145/1099554.1099689}
\showDOI{\tempurl}


\bibitem[Hamilton et~al\mbox{.}(2017)]%
        {hamilton_graph}
\bibfield{author}{\bibinfo{person}{William~L. Hamilton}, \bibinfo{person}{Rex
  Ying}, {and} \bibinfo{person}{Jure Leskovec}.}
  \bibinfo{year}{2017}\natexlab{}.
\newblock \showarticletitle{Inductive Representation Learning on Large Graphs}.
  In \bibinfo{booktitle}{\emph{Proceedings of the 31st International Conference
  on Neural Information Processing Systems}} (Long Beach, California, USA)
  \emph{(\bibinfo{series}{NIPS'17})}. \bibinfo{publisher}{Curran Associates
  Inc.}, \bibinfo{address}{Red Hook, NY, USA}, \bibinfo{pages}{1025–1035}.
\newblock
\showISBNx{9781510860964}


\bibitem[Hassan et~al\mbox{.}(2022)]%
        {exponential_2022}
\bibfield{author}{\bibinfo{person}{Ayat~Yehia Hassan}, \bibinfo{person}{Etimad
  Fadel}, {and} \bibinfo{person}{Nadine Akkari}.}
  \bibinfo{year}{2022}\natexlab{}.
\newblock \showarticletitle{Exponential {Decay} {Function}-{Based}
  {Time}-{Aware} {Recommender} {System} for e-{Commerce} {Applications}}.
\newblock \bibinfo{journal}{\emph{International Journal of Advanced Computer
  Science and Applications}} \bibinfo{volume}{13}, \bibinfo{number}{10}
  (\bibinfo{year}{2022}).
\newblock
\showISSN{21565570, 2158107X}
\urldef\tempurl%
\url{https://doi.org/10.14569/IJACSA.2022.0131071}
\showDOI{\tempurl}


\bibitem[He et~al\mbox{.}(2020)]%
        {LightGCN}
\bibfield{author}{\bibinfo{person}{Xiangnan He}, \bibinfo{person}{Kuan Deng},
  \bibinfo{person}{Xiang Wang}, \bibinfo{person}{Yan Li},
  \bibinfo{person}{YongDong Zhang}, {and} \bibinfo{person}{Meng Wang}.}
  \bibinfo{year}{2020}\natexlab{}.
\newblock \showarticletitle{LightGCN: Simplifying and Powering Graph
  Convolution Network for Recommendation}. In
  \bibinfo{booktitle}{\emph{Proceedings of the 43rd International ACM SIGIR
  Conference on Research and Development in Information Retrieval}} (Virtual
  Event, China) \emph{(\bibinfo{series}{SIGIR '20})}.
  \bibinfo{publisher}{Association for Computing Machinery},
  \bibinfo{address}{New York, NY, USA}, \bibinfo{pages}{639–648}.
\newblock
\showISBNx{9781450380164}
\urldef\tempurl%
\url{https://doi.org/10.1145/3397271.3401063}
\showDOI{\tempurl}


\bibitem[He et~al\mbox{.}(2018)]%
        {att_hist}
\bibfield{author}{\bibinfo{person}{Xiangnan He}, \bibinfo{person}{Zhankui He},
  \bibinfo{person}{Jingkuan Song}, \bibinfo{person}{Zhenguang Liu},
  \bibinfo{person}{Yu-Gang Jiang}, {and} \bibinfo{person}{Tat-Seng Chua}.}
  \bibinfo{year}{2018}\natexlab{}.
\newblock \showarticletitle{{NAIS}: Neural Attentive Item Similarity Model for
  Recommendation}.
\newblock \bibinfo{journal}{\emph{{IEEE} Transactions on Knowledge and Data
  Engineering}} \bibinfo{volume}{30}, \bibinfo{number}{12} (\bibinfo{date}{dec}
  \bibinfo{year}{2018}), \bibinfo{pages}{2354--2366}.
\newblock
\urldef\tempurl%
\url{https://doi.org/10.1109/tkde.2018.2831682}
\showDOI{\tempurl}


\bibitem[He et~al\mbox{.}(2017)]%
        {ncf}
\bibfield{author}{\bibinfo{person}{Xiangnan He}, \bibinfo{person}{Lizi Liao},
  \bibinfo{person}{Hanwang Zhang}, \bibinfo{person}{Liqiang Nie},
  \bibinfo{person}{Xia Hu}, {and} \bibinfo{person}{Tat-Seng Chua}.}
  \bibinfo{year}{2017}\natexlab{}.
\newblock \showarticletitle{Neural Collaborative Filtering}. In
  \bibinfo{booktitle}{\emph{Proceedings of the 26th International Conference on
  World Wide Web}} (Perth, Australia) \emph{(\bibinfo{series}{WWW '17})}.
  \bibinfo{publisher}{International World Wide Web Conferences Steering
  Committee}, \bibinfo{address}{Republic and Canton of Geneva, CHE},
  \bibinfo{pages}{173–182}.
\newblock
\showISBNx{9781450349130}
\urldef\tempurl%
\url{https://doi.org/10.1145/3038912.3052569}
\showDOI{\tempurl}


\bibitem[Hu et~al\mbox{.}(2008)]%
        {yohan}
\bibfield{author}{\bibinfo{person}{Yifan Hu}, \bibinfo{person}{Yehuda Koren},
  {and} \bibinfo{person}{Chris Volinsky}.} \bibinfo{year}{2008}\natexlab{}.
\newblock \showarticletitle{Collaborative Filtering for Implicit Feedback
  Datasets}. In \bibinfo{booktitle}{\emph{2008 Eighth IEEE International
  Conference on Data Mining}}. \bibinfo{pages}{263--272}.
\newblock
\urldef\tempurl%
\url{https://doi.org/10.1109/ICDM.2008.22}
\showDOI{\tempurl}


\bibitem[Kingma and Ba(2014)]%
        {kingma2014adam}
\bibfield{author}{\bibinfo{person}{Diederik~P Kingma} {and}
  \bibinfo{person}{Jimmy Ba}.} \bibinfo{year}{2014}\natexlab{}.
\newblock \showarticletitle{Adam: A method for stochastic optimization}.
\newblock \bibinfo{journal}{\emph{arXiv preprint arXiv:1412.6980}}
  (\bibinfo{year}{2014}).
\newblock


\bibitem[Koren(2008)]%
        {svdplus}
\bibfield{author}{\bibinfo{person}{Yehuda Koren}.}
  \bibinfo{year}{2008}\natexlab{}.
\newblock \showarticletitle{Factorization Meets the Neighborhood: A
  Multifaceted Collaborative Filtering Model}. In
  \bibinfo{booktitle}{\emph{Proceedings of the 14th ACM SIGKDD International
  Conference on Knowledge Discovery and Data Mining}} (Las Vegas, Nevada, USA)
  \emph{(\bibinfo{series}{KDD '08})}. \bibinfo{publisher}{Association for
  Computing Machinery}, \bibinfo{address}{New York, NY, USA},
  \bibinfo{pages}{426–434}.
\newblock
\showISBNx{9781605581934}
\urldef\tempurl%
\url{https://doi.org/10.1145/1401890.1401944}
\showDOI{\tempurl}


\bibitem[Koren(2009)]%
        {koren_bias}
\bibfield{author}{\bibinfo{person}{Yehuda Koren}.}
  \bibinfo{year}{2009}\natexlab{}.
\newblock \showarticletitle{Collaborative filtering with temporal dynamics}. In
  \bibinfo{booktitle}{\emph{Proceedings of the 15th {ACM} {SIGKDD}
  international conference on {Knowledge} discovery and data mining}}
  \emph{(\bibinfo{series}{{KDD} '09})}. \bibinfo{publisher}{Association for
  Computing Machinery}, \bibinfo{address}{New York, NY, USA},
  \bibinfo{pages}{447--456}.
\newblock
\showISBNx{978-1-60558-495-9}
\urldef\tempurl%
\url{https://doi.org/10.1145/1557019.1557072}
\showDOI{\tempurl}


\bibitem[Koren et~al\mbox{.}(2009)]%
        {mf}
\bibfield{author}{\bibinfo{person}{Yehuda Koren}, \bibinfo{person}{Robert
  Bell}, {and} \bibinfo{person}{Chris Volinsky}.}
  \bibinfo{year}{2009}\natexlab{}.
\newblock \showarticletitle{Matrix Factorization Techniques for Recommender
  Systems}.
\newblock \bibinfo{journal}{\emph{Computer}} \bibinfo{volume}{42},
  \bibinfo{number}{8} (\bibinfo{year}{2009}), \bibinfo{pages}{30--37}.
\newblock
\urldef\tempurl%
\url{https://doi.org/10.1109/MC.2009.263}
\showDOI{\tempurl}


\bibitem[Li et~al\mbox{.}(2020a)]%
        {TIASA_sequential_2019}
\bibfield{author}{\bibinfo{person}{Jiacheng Li}, \bibinfo{person}{Yujie Wang},
  {and} \bibinfo{person}{Julian McAuley}.} \bibinfo{year}{2020}\natexlab{a}.
\newblock \showarticletitle{Time {Interval} {Aware} {Self}-{Attention} for
  {Sequential} {Recommendation}}. In \bibinfo{booktitle}{\emph{Proceedings of
  the 13th {International} {Conference} on {Web} {Search} and {Data}
  {Mining}}}. \bibinfo{publisher}{ACM}, \bibinfo{address}{Houston TX USA},
  \bibinfo{pages}{322--330}.
\newblock
\showISBNx{978-1-4503-6822-3}
\urldef\tempurl%
\url{https://doi.org/10.1145/3336191.3371786}
\showDOI{\tempurl}


\bibitem[Li et~al\mbox{.}(2020b)]%
        {dgcf}
\bibfield{author}{\bibinfo{person}{X. Li}, \bibinfo{person}{M. Zhang},
  \bibinfo{person}{S. Wu}, \bibinfo{person}{Z. Liu}, \bibinfo{person}{L. Wang},
  {and} \bibinfo{person}{P.~S. Yu}.} \bibinfo{year}{2020}\natexlab{b}.
\newblock \showarticletitle{Dynamic Graph Collaborative Filtering}. In
  \bibinfo{booktitle}{\emph{2020 IEEE International Conference on Data Mining
  (ICDM)}}. \bibinfo{publisher}{IEEE Computer Society}, \bibinfo{address}{Los
  Alamitos, CA, USA}, \bibinfo{pages}{322--331}.
\newblock
\urldef\tempurl%
\url{https://doi.org/10.1109/ICDM50108.2020.00041}
\showDOI{\tempurl}


\bibitem[Liu et~al\mbox{.}(2020)]%
        {dtmf_2020}
\bibfield{author}{\bibinfo{person}{Tongtong Liu}, \bibinfo{person}{Wenming Ma},
  {and} \bibinfo{person}{Yulong Song}.} \bibinfo{year}{2020}\natexlab{}.
\newblock \showarticletitle{Deep {Time}-{Aware} {Matrix} {Factorization}}. In
  \bibinfo{booktitle}{\emph{2020 13th {International} {Congress} on {Image} and
  {Signal} {Processing}, {BioMedical} {Engineering} and {Informatics}
  ({CISP}-{BMEI})}}. \bibinfo{pages}{952--956}.
\newblock
\urldef\tempurl%
\url{https://doi.org/10.1109/CISP-BMEI51763.2020.9263503}
\showDOI{\tempurl}


\bibitem[Nasraoui et~al\mbox{.}(2007)]%
        {timewindow}
\bibfield{author}{\bibinfo{person}{O. Nasraoui}, \bibinfo{person}{Jeff
  Cerwinske}, \bibinfo{person}{Carlos Rojas}, {and} \bibinfo{person}{Fabio~A.
  Gonz{\'a}lez}.} \bibinfo{year}{2007}\natexlab{}.
\newblock \showarticletitle{Performance of Recommendation Systems in Dynamic
  Streaming Environments}. In \bibinfo{booktitle}{\emph{SDM}}.
\newblock


\bibitem[Rendle et~al\mbox{.}(2009)]%
        {bpr}
\bibfield{author}{\bibinfo{person}{Steffen Rendle}, \bibinfo{person}{Christoph
  Freudenthaler}, \bibinfo{person}{Zeno Gantner}, {and} \bibinfo{person}{Lars
  Schmidt-Thieme}.} \bibinfo{year}{2009}\natexlab{}.
\newblock \showarticletitle{BPR: Bayesian Personalized Ranking from Implicit
  Feedback}. In \bibinfo{booktitle}{\emph{Proceedings of the Twenty-Fifth
  Conference on Uncertainty in Artificial Intelligence}} (Montreal, Quebec,
  Canada) \emph{(\bibinfo{series}{UAI '09})}. \bibinfo{publisher}{AUAI Press},
  \bibinfo{address}{Arlington, Virginia, USA}, \bibinfo{pages}{452–461}.
\newblock
\showISBNx{9780974903958}


\bibitem[Tang and Wang(2018)]%
        {caser}
\bibfield{author}{\bibinfo{person}{Jiaxi Tang} {and} \bibinfo{person}{Ke
  Wang}.} \bibinfo{year}{2018}\natexlab{}.
\newblock \showarticletitle{Personalized Top-N Sequential Recommendation via
  Convolutional Sequence Embedding}. In \bibinfo{booktitle}{\emph{Proceedings
  of the Eleventh ACM International Conference on Web Search and Data Mining}}
  (Marina Del Rey, CA, USA) \emph{(\bibinfo{series}{WSDM '18})}.
  \bibinfo{publisher}{Association for Computing Machinery},
  \bibinfo{address}{New York, NY, USA}, \bibinfo{pages}{565–573}.
\newblock
\showISBNx{9781450355810}
\urldef\tempurl%
\url{https://doi.org/10.1145/3159652.3159656}
\showDOI{\tempurl}


\bibitem[Vaswani et~al\mbox{.}(2017)]%
        {attention}
\bibfield{author}{\bibinfo{person}{Ashish Vaswani}, \bibinfo{person}{Noam
  Shazeer}, \bibinfo{person}{Niki Parmar}, \bibinfo{person}{Jakob Uszkoreit},
  \bibinfo{person}{Llion Jones}, \bibinfo{person}{Aidan~N Gomez},
  \bibinfo{person}{\L~ukasz Kaiser}, {and} \bibinfo{person}{Illia Polosukhin}.}
  \bibinfo{year}{2017}\natexlab{}.
\newblock \showarticletitle{Attention is All you Need}. In
  \bibinfo{booktitle}{\emph{Advances in Neural Information Processing
  Systems}}, \bibfield{editor}{\bibinfo{person}{I.~Guyon},
  \bibinfo{person}{U.~Von Luxburg}, \bibinfo{person}{S.~Bengio},
  \bibinfo{person}{H.~Wallach}, \bibinfo{person}{R.~Fergus},
  \bibinfo{person}{S.~Vishwanathan}, {and} \bibinfo{person}{R.~Garnett}}
  (Eds.), Vol.~\bibinfo{volume}{30}. \bibinfo{publisher}{Curran Associates,
  Inc.}
\newblock
\urldef\tempurl%
\url{https://proceedings.neurips.cc/paper_files/paper/2017/file/3f5ee243547dee91fbd053c1c4a845aa-Paper.pdf}
\showURL{%
\tempurl}


\bibitem[Wang et~al\mbox{.}(2021)]%
        {survery_SRSs}
\bibfield{author}{\bibinfo{person}{Shoujin Wang}, \bibinfo{person}{Longbing
  Cao}, \bibinfo{person}{Yan Wang}, \bibinfo{person}{Quan~Z. Sheng},
  \bibinfo{person}{Mehmet~A. Orgun}, {and} \bibinfo{person}{Defu Lian}.}
  \bibinfo{year}{2021}\natexlab{}.
\newblock \showarticletitle{A Survey on Session-Based Recommender Systems}.
\newblock \bibinfo{journal}{\emph{ACM Comput. Surv.}} \bibinfo{volume}{54},
  \bibinfo{number}{7}, Article \bibinfo{articleno}{154} (\bibinfo{date}{jul}
  \bibinfo{year}{2021}), \bibinfo{numpages}{38}~pages.
\newblock
\showISSN{0360-0300}
\urldef\tempurl%
\url{https://doi.org/10.1145/3465401}
\showDOI{\tempurl}


\bibitem[Wang et~al\mbox{.}(2019b)]%
        {SRSs}
\bibfield{author}{\bibinfo{person}{Shoujin Wang}, \bibinfo{person}{Liang Hu},
  \bibinfo{person}{Yan Wang}, \bibinfo{person}{Longbing Cao},
  \bibinfo{person}{Quan~Z. Sheng}, {and} \bibinfo{person}{Mehmet Orgun}.}
  \bibinfo{year}{2019}\natexlab{b}.
\newblock \showarticletitle{Sequential {Recommender} {Systems}: {Challenges},
  {Progress} and {Prospects}}. In \bibinfo{booktitle}{\emph{Proceedings of the
  {Twenty}-{Eighth} {International} {Joint} {Conference} on {Artificial}
  {Intelligence}}}. \bibinfo{publisher}{International Joint Conferences on
  Artificial Intelligence Organization}, \bibinfo{address}{Macao, China},
  \bibinfo{pages}{6332--6338}.
\newblock
\showISBNx{978-0-9992411-4-1}
\urldef\tempurl%
\url{https://doi.org/10.24963/ijcai.2019/883}
\showDOI{\tempurl}


\bibitem[Wang et~al\mbox{.}(2019a)]%
        {wang_neural_2019}
\bibfield{author}{\bibinfo{person}{Xiang Wang}, \bibinfo{person}{Xiangnan He},
  \bibinfo{person}{Meng Wang}, \bibinfo{person}{Fuli Feng}, {and}
  \bibinfo{person}{Tat-Seng Chua}.} \bibinfo{year}{2019}\natexlab{a}.
\newblock \showarticletitle{Neural {Graph} {Collaborative} {Filtering}}. In
  \bibinfo{booktitle}{\emph{Proceedings of the 42nd {International} {ACM}
  {SIGIR} {Conference} on {Research} and {Development} in {Information}
  {Retrieval}}} \emph{(\bibinfo{series}{{SIGIR}'19})}.
  \bibinfo{publisher}{Association for Computing Machinery},
  \bibinfo{address}{New York, NY, USA}, \bibinfo{pages}{165--174}.
\newblock
\showISBNx{978-1-4503-6172-9}
\urldef\tempurl%
\url{https://doi.org/10.1145/3331184.3331267}
\showDOI{\tempurl}


\bibitem[Wu et~al\mbox{.}(2022)]%
        {GNNBook2022}
\bibfield{author}{\bibinfo{person}{Lingfei Wu}, \bibinfo{person}{Peng Cui},
  \bibinfo{person}{Jian Pei}, {and} \bibinfo{person}{Liang Zhao}.}
  \bibinfo{year}{2022}\natexlab{}.
\newblock \bibinfo{booktitle}{\emph{Graph Neural Networks: Foundations,
  Frontiers, and Applications}}.
\newblock \bibinfo{publisher}{Springer Singapore},
  \bibinfo{address}{Singapore}. 725 pages.
\newblock


\bibitem[Wu et~al\mbox{.}(2019)]%
        {sr_gnn}
\bibfield{author}{\bibinfo{person}{Shu Wu}, \bibinfo{person}{Yuyuan Tang},
  \bibinfo{person}{Yanqiao Zhu}, \bibinfo{person}{Liang Wang},
  \bibinfo{person}{Xing Xie}, {and} \bibinfo{person}{Tieniu Tan}.}
  \bibinfo{year}{2019}\natexlab{}.
\newblock \showarticletitle{{Session-based Recommendation with Graph Neural
  Networks}}. In \bibinfo{booktitle}{\emph{Proceedings of the Twenty-Third AAAI
  Conference on Artificial Intelligence}} (Honolulu, HI, USA),
  \bibfield{editor}{\bibinfo{person}{Pascal~Van Hentenryck} {and}
  \bibinfo{person}{Zhi-Hua Zhou}} (Eds.), Vol.~\bibinfo{volume}{33}.
  \bibinfo{publisher}{AAAI Press}, \bibinfo{pages}{346--353}.
\newblock
\urldef\tempurl%
\url{https://doi.org/10.1609/aaai.v33i01.3301346}
\showDOI{\tempurl}


\bibitem[Xiong et~al\mbox{.}(2010)]%
        {temporal_cf}
\bibfield{author}{\bibinfo{person}{Liang Xiong}, \bibinfo{person}{X. Chen},
  \bibinfo{person}{Tzu-Kuo Huang}, \bibinfo{person}{Jeff~G. Schneider}, {and}
  \bibinfo{person}{Jaime~G. Carbonell}.} \bibinfo{year}{2010}\natexlab{}.
\newblock \showarticletitle{Temporal Collaborative Filtering with Bayesian
  Probabilistic Tensor Factorization}. In \bibinfo{booktitle}{\emph{SDM}}.
\newblock


\bibitem[Xue et~al\mbox{.}(2017)]%
        {xue_deep_2017}
\bibfield{author}{\bibinfo{person}{Hong-Jian Xue}, \bibinfo{person}{Xinyu Dai},
  \bibinfo{person}{Jianbing Zhang}, \bibinfo{person}{Shujian Huang}, {and}
  \bibinfo{person}{Jiajun Chen}.} \bibinfo{year}{2017}\natexlab{}.
\newblock \showarticletitle{Deep {Matrix} {Factorization} {Models} for
  {Recommender} {Systems}}. In \bibinfo{booktitle}{\emph{Proceedings of the
  {Twenty}-{Sixth} {International} {Joint} {Conference} on {Artificial}
  {Intelligence}}}. \bibinfo{publisher}{International Joint Conferences on
  Artificial Intelligence Organization}, \bibinfo{address}{Melbourne,
  Australia}, \bibinfo{pages}{3203--3209}.
\newblock
\showISBNx{978-0-9992411-0-3}
\urldef\tempurl%
\url{https://doi.org/10.24963/ijcai.2017/447}
\showDOI{\tempurl}


\bibitem[Ying et~al\mbox{.}(2018)]%
        {pinsage}
\bibfield{author}{\bibinfo{person}{Rex Ying}, \bibinfo{person}{Ruining He},
  \bibinfo{person}{Kaifeng Chen}, \bibinfo{person}{Pong Eksombatchai},
  \bibinfo{person}{William~L. Hamilton}, {and} \bibinfo{person}{Jure
  Leskovec}.} \bibinfo{year}{2018}\natexlab{}.
\newblock \showarticletitle{Graph Convolutional Neural Networks for Web-Scale
  Recommender Systems}. In \bibinfo{booktitle}{\emph{Proceedings of the 24th
  {ACM} {SIGKDD} International Conference on Knowledge Discovery {\&} Data
  Mining, {KDD} 2018, London, UK, August 19-23, 2018}},
  \bibfield{editor}{\bibinfo{person}{Yike Guo} {and} \bibinfo{person}{Faisal
  Farooq}} (Eds.). \bibinfo{publisher}{{ACM}}, \bibinfo{pages}{974--983}.
\newblock
\urldef\tempurl%
\url{https://doi.org/10.1145/3219819.3219890}
\showDOI{\tempurl}


\bibitem[Zhu et~al\mbox{.}(2017)]%
        {rnn_fail}
\bibfield{author}{\bibinfo{person}{Yu Zhu}, \bibinfo{person}{Hao Li},
  \bibinfo{person}{Yikang Liao}, \bibinfo{person}{Beidou Wang},
  \bibinfo{person}{Ziyu Guan}, \bibinfo{person}{Haifeng Liu}, {and}
  \bibinfo{person}{Deng Cai}.} \bibinfo{year}{2017}\natexlab{}.
\newblock \showarticletitle{What to Do next: Modeling User Behaviors by
  Time-LSTM}. In \bibinfo{booktitle}{\emph{Proceedings of the 26th
  International Joint Conference on Artificial Intelligence}} (Melbourne,
  Australia) \emph{(\bibinfo{series}{IJCAI'17})}. \bibinfo{publisher}{AAAI
  Press}, \bibinfo{pages}{3602–3608}.
\newblock
\showISBNx{9780999241103}


\end{thebibliography}

\end{document}